\documentclass[preprint, preprintnumbers ,showpacs,amsmath,amssymb, superscriptaddress]{revtex4-1}
\pdfoutput=1
\usepackage{hyperref}
\usepackage{amsmath}
\usepackage{amssymb}
\usepackage{bm}
\usepackage{amsthm}
\usepackage{amsfonts}
\usepackage{latexsym}
\usepackage{graphicx}
\usepackage[usenames,dvipsnames]{color}
\usepackage{microtype}
\usepackage{pdfpages}
\usepackage{ifpdf}

\usepackage[normalem]{ulem}
\begin{document}

\title{
Heterointerface engineered electronic and magnetic phases of a correlated complex oxide
}
\author{Jian~Liu} %\email{jxl026@uark.edu}
\affiliation{Department of Physics, University of Arkansas, Fayetteville, Arkansas 72701, USA}
\affiliation{Advanced Light Source, Lawrence Berkeley National Laboratory, Berkeley, California 94720, USA}
\affiliation{Materials Science Division, Lawrence Berkeley National Laboratory, Berkeley, California 94720, USA}
\author{Mehdi Kargarian}
\affiliation{Department of Physics, The University of Texas at Austin, Austin, Texas 78712, USA}
\author{Mikhail Kareev}
\affiliation{Department of Physics, University of Arkansas, Fayetteville, Arkansas 72701, USA}
\author{Ben Gray}
\affiliation{Department of Physics, University of Arkansas, Fayetteville, Arkansas 72701, USA}
\author{Phil Ryan}
\affiliation{Advanced Photon Source, Argonne National Laboratory, Argonne, Illinois 60439, USA}
\author{Alejandro Cruz}
\affiliation{Advanced Light Source, Lawrence Berkeley National Laboratory, Berkeley, California 94720, USA}
\author{Nadeem Tahir}
\affiliation{Advanced Light Source, Lawrence Berkeley National Laboratory, Berkeley, California 94720, USA}
\author{Yi-De Chuang}
\affiliation{Advanced Light Source, Lawrence Berkeley National Laboratory, Berkeley, California 94720, USA}
\author{Jing-Hua Guo}
\affiliation{Advanced Light Source, Lawrence Berkeley National Laboratory, Berkeley, California 94720, USA}
\author{James M. Rondinelli}
\affiliation{Department of Materials Science and Engineering, Drexel University,  Philadelphia, Pennsylvania 19104, USA}
\author{John W. Freeland}
\affiliation{Advanced Photon Source, Argonne National Laboratory, Argonne, Illinois 60439, USA}
\author{Gregory A. Fiete}
\affiliation{Department of Physics, The University of Texas at Austin, Austin, Texas 78712, USA}
\author{Jak Chakhalian}
\affiliation{Department of Physics, University of Arkansas, Fayetteville, Arkansas 72701, USA}

%\date{\today}
\maketitle
\sloppy

%\section{Introduction}

{\bf The physics of Mott systems is characterized by an interaction-driven metal-to-insulator transition (MIT) in a partially filled band.  In the resulting insulating state, magnetic order of the local moments (often antiferromagnetic (AFM)) typically develops, but in rare situations no long-range magnetic order appears, even at zero temperature, rendering the system a ``quantum spin liquid".  In Mott insulating oxides, intriguing charge, spin and/or orbital orderings are often found in the presence of localized carriers, e.g. in manganites, while mobile carriers may experience strong quantum fluctuations resulting in non-Fermi liquid (NFL) behavior, such as the ``strange metal regime" (a linear-$T$ resistivity) in the cuprates.
Despite this diversity, the underlying energetic landscape in these materials is derived from three essential parameters: the on-site electron-electron repulsion $U$, the energy difference $\Delta$ from the empty local $d$-orbitals to the band-like oxygen $p$-states, and their hybridization strength, which is encapsulated in the bandwidth, $W$.  %The ratios $U/W$ and $W/\Delta$ describe the degrees of correlation and covalency, respectively.
A fundamental and technologically critical question is whether one can tune these parameters to control both MITs and Neel transitions, and even stabilize latent metastable phases, ideally on a platform suitable for applications.
Here we demonstrate how to achieve control of all these features in ultrathin films of NdNiO$_{3}$ grown on substrates of various degrees of lattice mismatch. In particular, upon the decay of the AFM Mott insulating state into a stable NFL phase distinctly different from that in the cuprates, we find evidence of a quantum MIT that spans a non-magnetic insulating phase (possibly a quantum spin liquid). These quantum critical behaviors are not observed in the bulk phase diagram of NdNiO$_{3}$.}

%eveals the melting of the AFM Mott insulating state into a quantum critical non-Fermi liquid (NFL) with an intermediate spin-disordered state that is not accessible in the bulk. This emergent phase behavior induced by epitaxial strain is realized through the simultaneous modulation of covalency and correlation via the Madelung cohesive energy and ligand hole density, indicated by soft x-ray spectroscopy.}

\begin{figure*}[t]\vspace{-0pt}
\includegraphics[width=\textwidth]{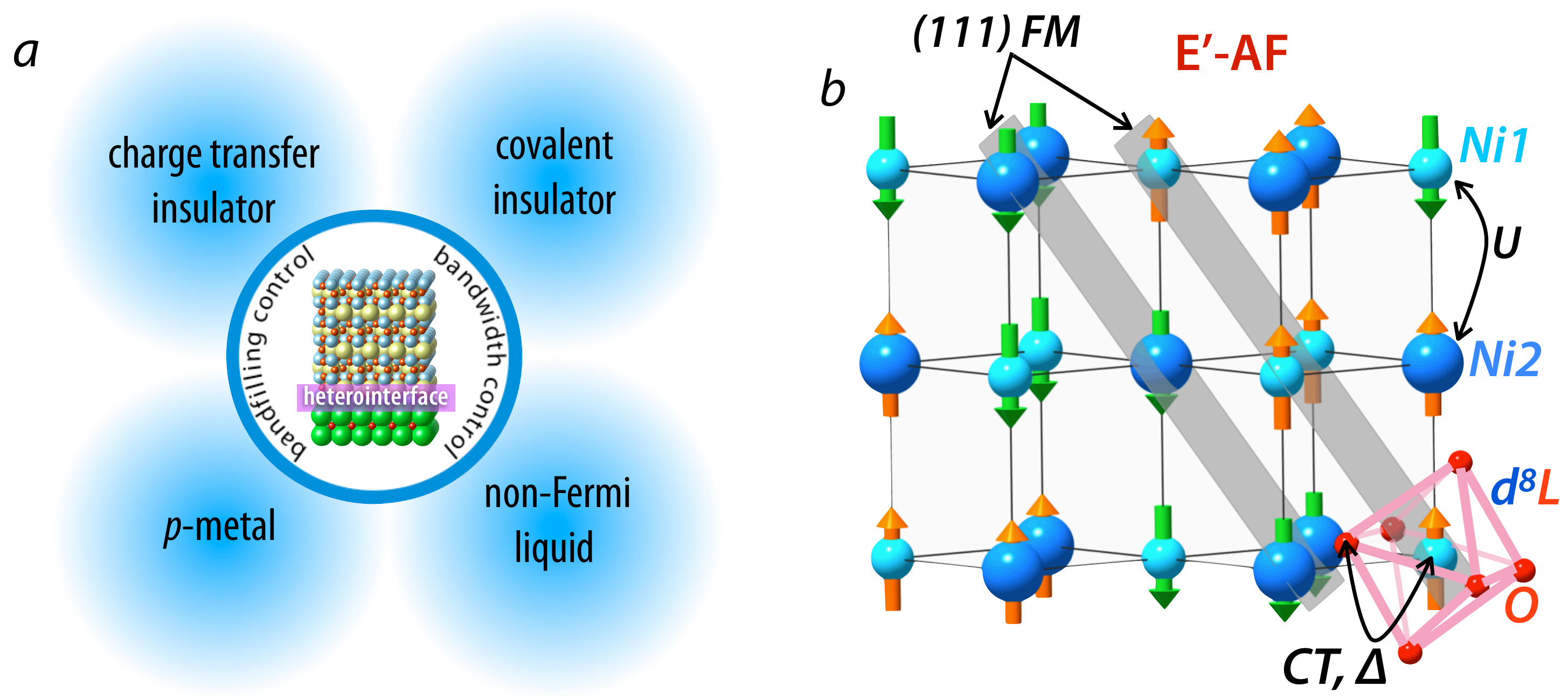}
\caption{\label{phase-control} (a) Accessibility of latent phases in the heterointerface system.  Interfacial constraints control both the band filling through a `self-doping' effect and the bandwidth allowing new phases to appear that are not observed in bulk systems, even under significant pressure and chemical doping.  Phases that may be obtained include a charge-transfer insulator, covalent metal, $p$-metal, and non-Fermi liquid.  In this work, we report an emergent paramagnetic insulating regime and a remarkably stable non-Fermi liquid over a significant portion of the phase diagram. (b) Four NdNiO$_{3}$ perovskite unit cells. Arrows denote spins in the associated E$^{\prime}$--type AFM structure with the $\uparrow-\uparrow-\downarrow-\downarrow$ stacking of the pseudo-cubic (111) FM planes, two of which are highlighted in gray. $L$ denotes an oxygen hole.
}
\end{figure*}

With recent advances in the atomic layering of correlated oxides \cite{Chakhalian2012,Mannhart}, a new route has been established for manipulating the low-energy electronic structure at the nanoscale. Although the quantum criticality of the MIT and the associated magnetic transition has been a key issue in correlated electron systems for decades, it has not been investigated under the versatile controls of oxide heteroepitaxy due to the formidable challenges of probing AFM order in ultrathin layers.  Leveraging this experimental approach with theoretical advances, however, could bring us closer to ultimately establishing general ``design" rules for engineering desired phases in complex oxide heterostructures (Fig.~\ref{phase-control}(a)).
With this goal in mind, we performed a detailed study on ultrathin  $\sim\!5.7$ nm (15 unit cells) films of fully strained NdNiO$_{3}$ (oriented along the pseudocubic [001]-direction) synthesized by laser Molecular Beam Epitaxy (MBE) in a layer-by-layer fashion as described in Ref.[\onlinecite{Jian}].  A series of high-quality perovskite-based single crystal substrates is used to attain a wide range of lattice mismatch $\varepsilon$ from $-2.9$\% to $+4$\% (more details in \cite{Supplemental}). The results reveal full control of epitaxy on the MIT and the magnetic ordering. Specifically, tuning the amount of epitaxial strain from the tensile to compressive side first merges the MIT and AFM transition, followed by a rapid decay of the magnetic ordering into a spin-disordered phase (possible quantum spin liquid) before stabilizing a conducting NFL phase. In this work, we show the underlying electronic reconstruction is associated with simultaneous modulation of the bandwidth, $W$, and the self-doping, determined by $\Delta$.

\begin{figure*}\vspace{-0pt}
\includegraphics[width=\textwidth]{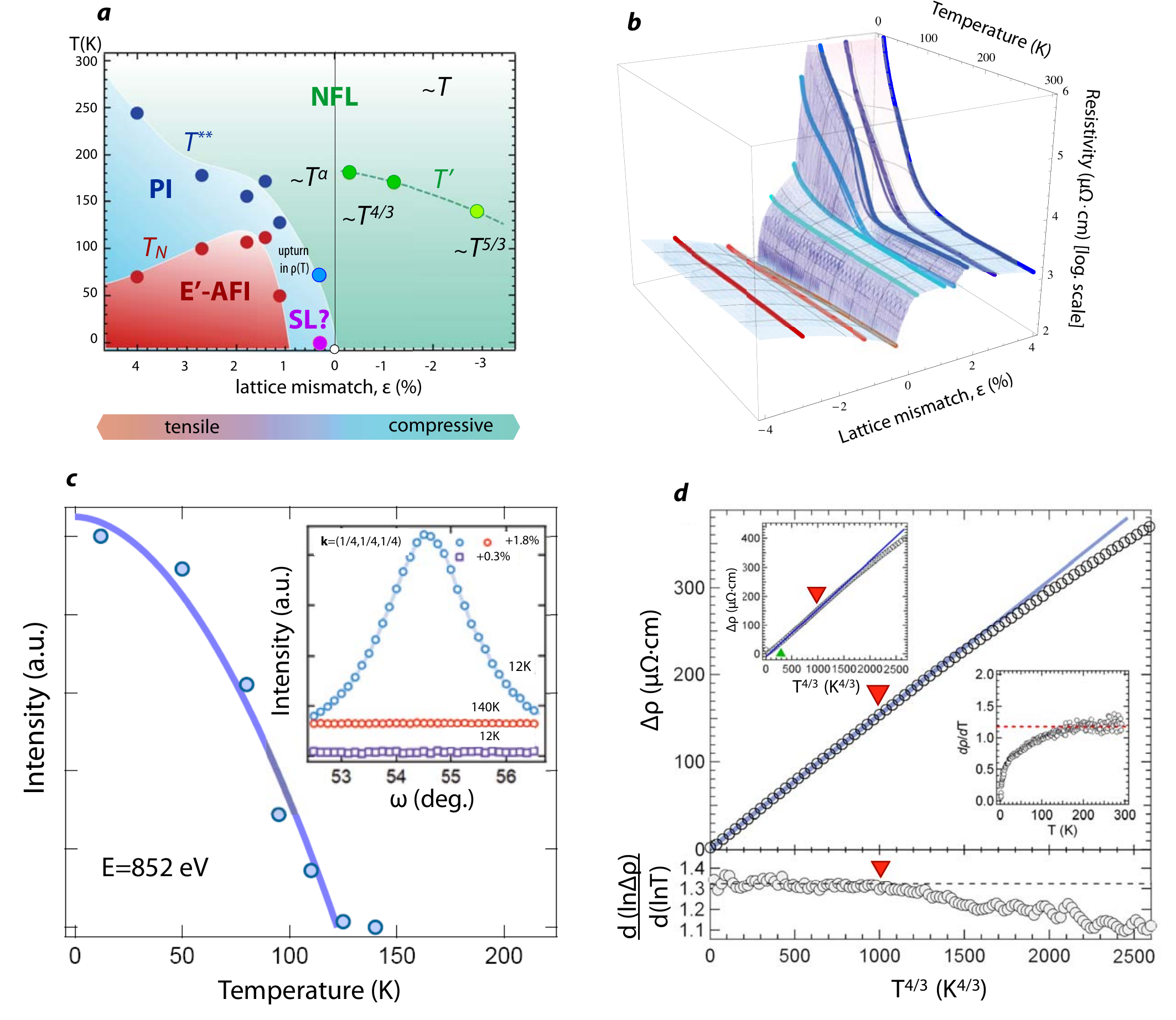}
\caption{\label{phase-diagram} (a) Lattice mismatch-temperature phase diagram. NFL, PI and AFI denote non-Fermi liquid, paramagnetic insulator and antiferromagnetic insulator, respectively. (b) Normalized temperature-dependent resistivity vs heteroepitaxial lattice mismatch. (c) Temperature dependence of the $E^{\prime}$-type AFM reflection intensity at $k=$(1/4,1/4,1/4) ((1/2,0,1/2) in orthorhombic notation) observed by resonant x-ray diffraction at the Ni $L_3$-edge and $\varepsilon=+1.8\%$. Inset: $\omega$ scans across the diffraction peak at 12 K and 140 K. (d) Fit-free analysis of the $T^{4/3}$ power-law behavior for $\varepsilon=-0.3\%$ and -1.2$\%$ (upper left inset). The lower right inset shows the temperature derivative of the resistivity, indicating a $T$-linear behavior at high temperatures. The upper red triangle signals the upper temperature limit $T'$ where the resistivity crosses over to a linear-$T$ behavior. The lower green triangle indicates the lower temperature where the temperature-dependence starts to deviate.}
\end{figure*}

\begin{figure*}\vspace{-0pt}
\includegraphics[width=\textwidth]{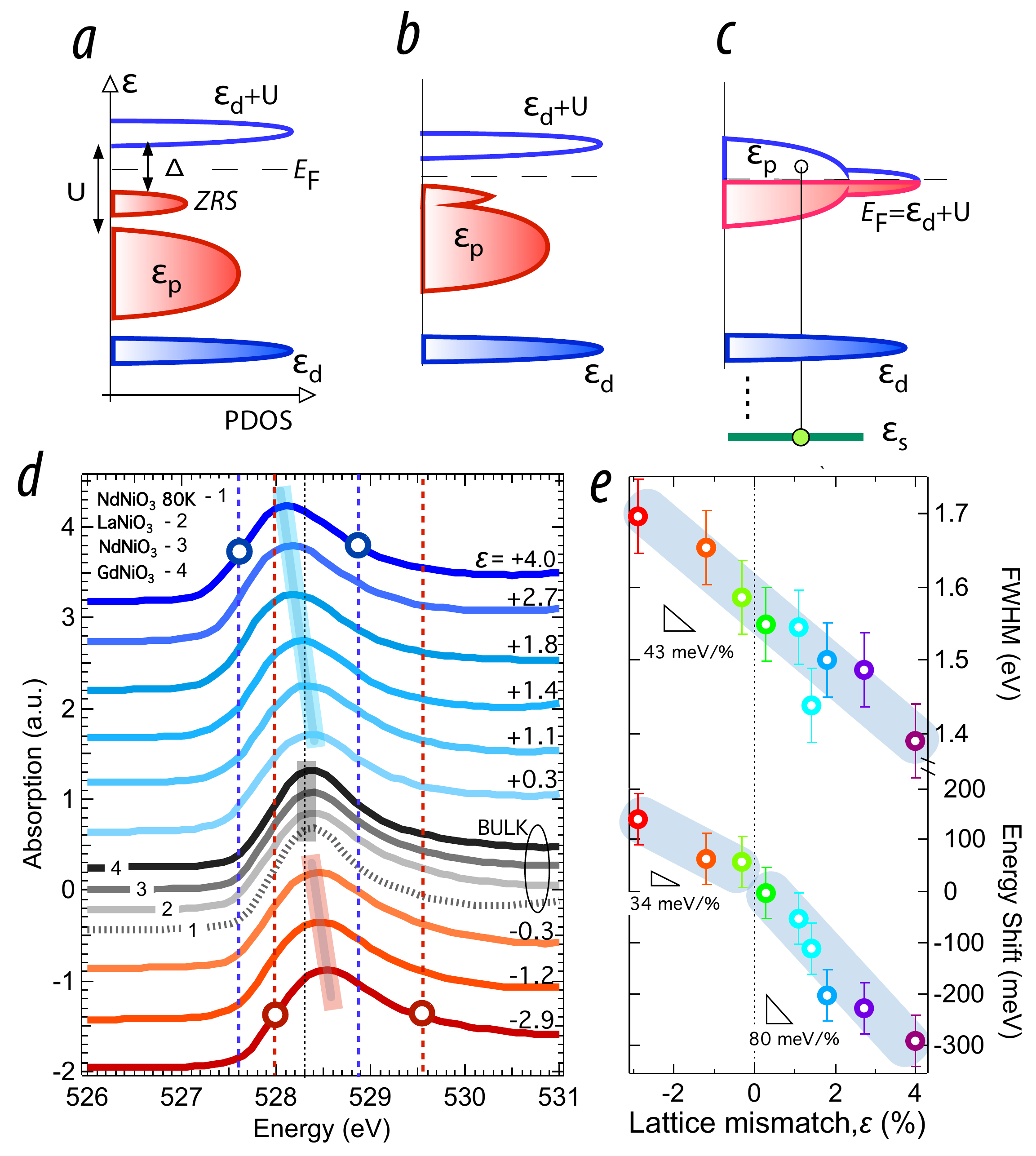}
\caption{\label{K-edge} (a)-(c) Schematics of partial density of states (PDOS) for charge transfer Mott materials with $d$ and $p$-type electrons. $\Delta$ is the charge-transfer energy, and $U$ is the on-site $d$-electron Coulomb repulsion energy. The core level and $1s\rightarrow2p$ transition involved in the O K-edge x-ray absorption are indicated in (c). (d) Soft x-ray absorption spectra of NdNiO$_3$ ultrathin films at O K-edge at 300 K. Three reference spectra at 300 K and one at 80 K from bulk LaNiO$_3$, NdNiO$_3$ and GdNiO$_3$ are included for comparison. Dashed lines are guides for eye. The open circles on the top and bottom spectra indicate the FWHM. (e) The pre-peak width FWHM around 529 eV and the energy shift of the pre-peak versus lattice mismatch $\varepsilon$.}
\end{figure*}

In the bulk, NdNiO$_{3}$ belongs to the charge-transfer nickelate family RENiO$_{3}$, where RE=Nd{\dots}Lu (except for La) are paramagnetic metals (PM) at high temperatures, but become insulating with charge-ordering and antiferromagnetically ordered at $T_{\textrm{MI}}$ and $T_\textrm{N}$, respectively \cite{Catalan0,Medarde1}. An important character of the AFM ordering is the presence of thermal hysteresis in transport properties around the transition due to the coupling between the spin and charge degrees of freedom when $T_{\textrm{N}}$ and $T_{\textrm{MI}}$ approaches each other, such as for RE=Pr and Nd in the bulk; as the  temperature is lowered well below $T_{\rm N}$  the transport hysteretic behavior is strongly suppressed and eventually  disappears \cite{Catalan0,Medarde1}.  Within the Sawatzky-Allen-Zaanen (SAZ) scheme, RENiO$_{3}$ belongs to  the class of  charge-transfer-type materials where the charge gap corresponds to the excitation of an oxygen 2$p$-electron into the unoccupied upper nickel $d$-band \cite{Zaanen,Mizokawa0,Sarma} as shown in Fig.~\ref{K-edge}(a)-(c).  The unusual high 3+ oxidation state of Ni and the  presence of a small excitation  energy ($\Delta\lesssim 1$ eV), as schematically  illustrated in  Fig.~\ref{K-edge}(b) and (c), naturally facilitates the transfer of oxygen $p$-electrons into the unoccupied nickel $d$-electron states  (or alternatively a transfer of a correlated hole onto oxygen) \cite{Khomskii}. This `self-doping' phenomenon results from the coupling of a band-like continuum of oxygen-derived states and localized correlated $d$-states, and is believed to be responsible for the unusual AFM spin ordering ($E^{\prime}-$type) \cite{Mizokawa,Catalan0,Medarde1,Alonso,Staub, Goodenough}, sometimes described as an ``$\uparrow-\uparrow-\downarrow-\downarrow$" stacking sequence of ferromagnetically (FM) ordered planes along the pseudo-cubic (111) direction which is characterized by the magnetic vector $\bf{k}=$(1/4,1/4,1/4) in cubic notation (see Fig.~\ref{phase-diagram}(d)).

%\subsection{Transport}

Fig.~\ref{phase-diagram}(a) summarizes the  evolution of the electronic and magnetic states in the lattice mismatch-temperature phase space.  As seen, despite the ultrathin form and the large lattice mismatch on some of the substrates, the metallicity is well preserved for all samples at room temperature. A direct inspection of the temperature-dependent resistivity curves from 5 to 300 K for different lattice mismatch, $\varepsilon$  in Fig.~\ref{phase-diagram}(b) indicates  the presence of well controlled and diverse electronic phase behaviors in the ultrathin films that are  absent in bulk NdNiO$_3$. Specifically, for samples in the positive $\varepsilon$ range (under tensile strain), the insulating ground state continuously develops with increasing magnitude of $\varepsilon$, whereas an unusual metallic NFL phase emerges and persists throughout the whole range of negative $\varepsilon$ (compressive strain).
The absence of the bulk-like first-order MIT near $\varepsilon \approx 0$ manifests the unique role of the heteroepitaxy in destabilization of Mott insulating state with charge and spin orderings; due to the interface-imposed lattice boundary condition, collective long-range order that strongly couples to the lattice degrees of freedom may be frustrated. In addition, even for $\varepsilon=0$, a substrate can still strongly distort the film structure via internal structural mismatches such as octahedral rotation, distortion and crystal symmetry.
Meanwhile, the strain-induced MIT signals that the heteroepitaxial NdNiO$_3$ is in close proximity to quantum criticality near $\varepsilon \approx 0$. Although the discrete values of $\varepsilon$ limit the ability to precisely pinpoint the location of the critical end point of the ``E$^{\prime}$-AFI" region, magnetism is rapidly suppressed on the insulating side upon approach to $\varepsilon \approx 0$ and appears to join the low-temperature NFL region for $\varepsilon<0$ by an intervening ground state without long-range magnetic ordering (a possible quantum spin liquid) as shown in Fig.~\ref{phase-diagram}(a). The intervening ground state is inferred at $\varepsilon=+0.3\%$ from the lack of thermal hysteresis in resistivity (which is present at all larger $\varepsilon$ values--see supplemental), and also the absence of the magnetic ordering peak in resonant X-ray diffraction (see below).

On the insulating side ($\varepsilon>0$), Fig.~\ref{phase-diagram}(a) shows the evolution of characteristic transition temperatures, from the high-temperature metallic phase to the intermediate-temperature paramagnetic insulating (PI) phase, $T^{**}$ (resistivity minimum temperature), and to the low-temperature $E'$-AFI  phase transition temperature, $T_\textrm{N}$ (the thermal hysteric inflection point of resistivity \cite{Supplemental}). In particular, for large values of $\varepsilon$ the $T^{**}$ is significantly higher than $T_\textrm{N}$. This behavior is  sharply distinct from  bulk NdNiO$_3$  where $T^{**}(=T_{\rm MI})$ coincides with $T_\textrm{N}$.  The resulting intermediate-temperature PI phase thus implies the opening of a gap which is decoupled from the spin ordering.  As $\varepsilon$ is reduced, $T^{**}$ and $T_\textrm{N}$ merge albeit with a different slope, $i.e.$  while $T^{**}$ quickly decreases, $T_\textrm{N}$ steadily rises until $\varepsilon \approx +1.8\%$.  This convergence of critical temperatures is further  evidenced in the enhanced thermal  hysteresis around $\varepsilon\approx +1.8\%$ [see Fig.~\ref{phase-diagram}(b) and Supplemental].  It is important to note that this PI phase  is  unattainable in bulk NdNiO$_3$ -- an example of a latent electronic phase in this system stabilized by the heterointerface.
As shown in Fig.~\ref{phase-diagram}(a), for $\varepsilon\lesssim+1.8\%$, the evolution of $T_\textrm{N}$ qualitatively changes so that  $T_\textrm{N}$ and $T^{**}$ exhibit approximately a ``parallel dive" in response to reducing $\varepsilon$. Upon further lowering $\varepsilon$ toward zero, $T_\textrm{N}$ rapidly vanishes  accompanied by a drastically weakened thermal hysteresis in resistivity \cite{Supplemental}, e.g. at $\varepsilon\approx +1.1\%$ (see Fig.~\ref{phase-diagram}(b)); the thermal hysteresis completely recedes from the system at around $\varepsilon\approx +0.3\%$ (see Supplemental for more detailed plots).  On the other hand, $T^{**}$ remains finite with the tendency  of suppression toward zero. This is in sharp contrast with the bulk where increasing hydrostatic pressure always favors magnetic ordering \cite{Zhou0}. %due to the bandwidth $W$-dependence of the exchange coupling, maintaining a strong first-order phase boundary while suppressing the MIT.
The difference in the behavior of $T^{**}$ and $T_\textrm{N}$ thus strongly suggests the emergence of an unusual weakly insulating ground state with completely quenched long-range AFM order in the vicinity of $\varepsilon=0\%$ .
To further corroborate this magnetic behavior, resonant magnetic X-ray diffraction \cite{Doering} at the Ni $L_3$-edge is utilized to directly track the $E^{\prime}-$type AFM ordering \cite{Bodenthin}. This measurement is done by monitoring the emerging intensity of the magnetic Bragg reflection at the magnetic vector $\bf{k}=$(1/4,1/4,1/4)  as a function of temperature, such as that at $\varepsilon=+1.8\%$ shown in Fig.~\ref{phase-diagram}(c). While the appearance of the magnetic peak at low temperatures is consistent with the $T_\textrm{N}$ extracted from the resistivity, no $E^{\prime}-$type AFM reflection is observed at $\varepsilon=+0.3\%$ down to 12 K (see the inset of Fig.~\ref{phase-diagram}(c)). The stabilization of this emergent spin-disordered state implies enhanced frustration in proximity to the zero temperature MIT, which presents an intriguing candidate for a quantum spin liquid (SL) \cite{Suter,Balents}.

Upon crossing the zero temperature MIT towards negative values of $\varepsilon$, the temperature-driven MIT is completely quenched and a new exotic metallic ground state emerges across the entire range of $\varepsilon <0$. To  stress the peculiarity  of  the phase we point out that at  ``high"  temperatures but still well below the Debye temperature $\sim$ 420 K\cite{Debye},  the  resistivity exhibits extended unconventional  linear $T$-dependence (see the bottom right inset of Fig.~\ref{phase-diagram}(d)) commonly seen in the ``strange metal" regime of the high-$T_{\rm c}$ cuprates \cite{Imada}, while Fermi liquids have a $T^{2}$-dependence. Upon crossing  the intermediate temperature scale ($\sim$150 K) marked as T$^{\prime}$ in Fig.~\ref{phase-diagram}(a), however,  another characteristic temperature dependence clearly appears. A fitting-free resistivity data analysis (see Fig.~\ref{phase-diagram}(d))  reveals a $T^{4/3}$ power-law behavior lingering over a 100 K temperature range.  The 4/3 power law behavior is characteristic of a NFL in the vicinity of a two-dimensional quantum critical point with dynamical exponent $z=3$ \cite{Maslov}.  For the  large negative values of $\varepsilon\lesssim   -2.9\%$,  the power of the NFL exponent below T$^\prime$ switches to 5/3 with increasing compressive strain \cite{Supplemental}.  The 5/3 exponent is characteristic of a three-dimensional critical point with dynamical exponent $z=3$ \cite{Maslov}. While we do not detect any sizable structural transition which might cause a change in the effective dimensionality of our system (from two-dimensional to three-dimensional) near $\varepsilon \approx -3\%$, theoretically, large bi-axial compression could drive such a transition in NdNiO$_3$\cite{Angel}.

%We note that a NFL with a 4/3 or 5/3 exponent has been reported for a different  bulk ceramic nickelate, PrNiO$_3$, subjected to high hydrostatic pressure \cite{Zhou0}.  A comparison with our system, however, reveals  that the heterointerface driven-NFL phase is characterized by a markedly  higher onset temperature T$^\prime$ ({\it i.e.} 200 K vs. 40 K in the bulk PrNiO$_3$).

The observed NFL features in transport  imply the  presence of strong quantum fluctuations stabilized by the hetero-epitaxial boundary. Indeed, as discussed above, the simultaneous rapid collapse of the AFM order and the emerging spin-disordered phase also highlights the important difference of this new quantum melting regime from that of the bulk where a $T_{\textrm{MI}}=T_\textrm{N}$ phase boundary is driven to a single critical end point.  To the best of our knowledge bulk  NdNiO$_3$ does not exhibit the NFL behavior reported here for $\varepsilon <0$ under  either hydrostatic or chemical pressure \cite{Catalan0,Medarde1}, nor are the 4/3 and 5/3 exponents commonly found across the RENiO$_3$ series.

It is remarkable that the NFL exponent is stabilized over such a wide range of temperatures and negative $\varepsilon$ in our experiments.  This behavior is qualitatively and even semi-quantitatively consistent with a Boltzmann-type transport theory \cite{Supplemental} involving multiple bands of different effective masses and zero-momentum critical fluctuations in the heaviest of the bands \cite{Maslov}.  Our density functional theory calculations   supports the multiple-band picture of different masses, and the structure of the transport theory\cite{Supplemental} provides  a natural mechanism for the crossover of a fractional exponent at lower temperatures (4/3 or 5/3) to the linear-$T$ behavior observed above T$^\prime$ but still below the Debye temperature.  The precise character of the zero-momentum quantum fluctuations remains unclear at present, and further experimental and theoretical work is due.

%\subsection{X-ray absorption}
In order  to  elucidate the electronic energy scales involved in controlling these emergent phases (PI, NFL, and possible quantum spin liquid), we have performed extensive resonant soft X-ray measurements  (XAS) on the oxygen K-edge, which $directly$ probes the hole state in the unoccupied $2p$-projected density of states \cite{Sarma02}. By utilizing the 1$s\rightarrow 2p$ transition on the oxygen K-edge (see Fig.~\ref{K-edge}(c)), i.e.\ $3d^{8}\underline{L}\rightarrow\underline{1s}3d^{8}$ ($\underline{L}$ denotes the ligand hole state), we evaluate the connection between the insulating phase behavior  and  `self-doping' behavior.  In addition, a set of  high-quality  bulk ceramic samples of LaNiO$_3$, NdNiO$_3$, and GdNiO$_3$ have been measured to  provide a benchmark comparison for resolving the underlying physics of heterointerface-control.

Figure~\ref{K-edge}(d) shows the resulting X-ray absorption  spectra obtained at the threshold energy around 528.5 eV, where the absorption pre-peak is exclusively due to Ni $3d$ states hybridized with O $2p$ states \cite{Sarma02} (also see \cite{Supplemental} for representative spectra in a wider energy range).
As clearly seen,  the pre-peak around 529 eV exhibits a remarkably  large and  asymmetric (with the sign of $\varepsilon$)  energy shift, indicative of an evolution in the charge excitation energy. Figure~\ref{K-edge}(e) quantifies the finding as follows: the oxygen-derived band edge moves downwards by as much as 270 meV at $\varepsilon$ = +4 (or $\sim$80(13) meV/\%) and upwards by $\sim$150 meV at $\varepsilon= -2.9$ (or $\sim$34(13) meV/\%). In sharp contrast to  this result, the shift is completely absent for the bulk data when  varying  chemical pressure and/or temperature crossing the  MIT into the charge-ordered AFM insulating state [shown as grey shaded curves in Fig.~\ref{K-edge}(d)].  These results point to the pivotal role of the epitaxial substrate lattice mismatch as the driving force for the observed shift. Although the decrease of the absorption threshold is strikingly similar to that seen from the introduction of holes and in-gap states through conventional chemical doping \cite{Merz}, the `hole doping' response here is achieved by shifting the entire prepeak in virtue of  heterointerface strain in the absence of explicit chemical doping \cite{footnote}.

The large observed shift of the excitation energy to the unoccupied states is connected to the shift of the O $1s$ core-level states with respect to the Ni $3d$-hybridized state.  This manifests itself through an altered relative Madelung site potential between Ni and O, which is the primary effect that defines the magnitude of the charge excitation energy $\Delta$ \cite{Ohta}. This finding lends strong support to  the  notion of a modulation of the fundament energy  scale -- $\Delta$ with  $\varepsilon$ \cite{Imada,Zaanen1}. Additionally, the pre-peak width, a measure of a degree of $p$-$d$ hybridization or covalency $W$,  scales almost  linearly  with $\varepsilon$ (see Fig.~\ref{K-edge}(d) and (e)), in accordance with the induced MIT. The combined modulations in both $\Delta$ and $W$ reflects the unique control of ultrathin NdNiO$_3$ in a `self-doped' material. In particular, the simultaneous regulation of the self-doped oxygen hole density via both $\Delta$ and $W$ is expected to tune the balance between the ferromagnetic and AFM exchange channels of the Ni-O-Ni bond in the $E^{\prime}-$type spin ordering \cite{Mizokawa}. Thus, deviation in the degree of self-doping would transpire to cause strong frustration and act to suppress the spin order, especially near the MIT, resulting in the collapse of the AFM ordering and a possible quantum spin liquid state.

%This observation bears intriguing resemblance to the rapidly quenched AFM order and the magnetically disordered insulating (and charge nematic) ground state found in underdoped cuprates \cite{Imada, Chou}.

In summary,  we have demonstrated the consequences of heterointerface constraints from substrate lattice mismatch and used it to drive emergent phase behavior and induce quantum critical behavior not accessible in the bulk series of RENiO$_3$.  This control is achieved through the modulation of the covalency,  $W$, and charge transfer energy, $\Delta$ with $\varepsilon$.  We have demonstrated that a specific ground state can be selected by  the fine balance of  the self-doped hole density on the oxygen atoms. We expect the physics uncovered for NdNiO$_3$ is rather general and should open the door to the rational design of new classes of correlated electron materials with a wider range of applications through an enriched phase diagram.

The authors acknowledge numerous insightful discussions with D. I. Khomskii, A. J. Millis, S. Okamoto and G. A. Sawatzky. J.C. was supported by DOD-ARO under the grant No. 0402-17291 and NSF grant No. DMR-0747808, M.K. and G.A.F. by DOD-ARO grant No. W911NF-09-1-0527,  W911NF-12-1-0573, and NSF Grant No. DMR-0955778. J.M.R. supported by DARPA under award no. N66001-12-1-4224. The Advanced Light Source is supported by the Director, Office of Science, Office of Basic Energy Sciences, of the U.S. Department of Energy under Contract No. DE-AC02-05CH11231. Work at the APS, Argonne is supported by the U.S. Department of Energy, Office of Science under grant No. DEAC02-06CH11357.

\pagebreak

\newpage
\newpage

\clearpage

\newpage


\section*{Supplementary material}

# Supplement for heterointerface engineered electronic and magnetic phases of a correlated complex oxide

Jian Liu,[1,2,3] Mehdi Kargarian,[4] Mikhail Kareev,[1] Ben Gray,[1] Phil Ryan,[5] Alejandro Cruz,[2] Nadeem Tahir,[2] Yi-De Chuang,[2] Jing-Hua Guo,[2] James M. Rondinelli,[6] John W. Freeland,[5] Gregory A. Fiete,[4] and Jak Chakhalian[1]

[1]Department of Physics, University of Arkansas, Fayetteville, Arkansas 72701, USA
[2]Advanced Light Source, Lawrence Berkeley National Laboratory, Berkeley, California 94720, USA
[3]Materials Science Division, Lawrence Berkeley National Laboratory, Berkeley, California 94720, USA
[4]Department of Physics, The University of Texas at Austin, Austin, Texas 78712, USA
[5]Advanced Photon Source, Argonne National Laboratory, Argonne, Illinois 60439, USA
[6]Department of Materials Science and Engineering,
Drexel University, Philadelphia, Pennsylvania 19104, USA

## I. EXPERIMENTAL DETAILS: SUBSTRATES, ELECTRICAL TRANSPORT, AND MAGNETIC MEASUREMENTS

Table I describes the set of synthetic single crystal substrates cut in pseudo-cubic (100) direction which are used for NdNiO$_3$ growth to facilitate controlled lattice mismatch, $\varepsilon$.

**TABLE I** | In-plane pseudo-cubic lattice constants, $a_{\text{sub}}$, of the used substrates and their corresponding lattice mismatch ($\varepsilon$) with NdNiO$_3$ (3.803 Å).

|  | YAlO$_3$ | SrLaAlO$_4$ | LaAlO$_3$ | SrPrGaO$_4$ | SrLaGaO$_4$ | NdGaO$_3$ | (La$_{0.3}$Sr$_{0.7}$)(Al$_{0.65}$Ta$_{0.35}$)O$_3$ | SrTiO$_3$ | DyScO$_3$ |
|---|---|---|---|---|---|---|---|---|---|
| $a_{sub}$ (Å) | 3.692 | 3.757 | 3.794 | 3.812 | 3.843 | 3.858 | 3.868 | 3.905 | 3.955 |
| $\varepsilon$ (%) | -2.9 | -1.2 | -0.3 | 0.3 | 1.1 | 1.4 | 1.8 | 2.7 | 4.0 |

Transport data are shown in Fig.1. Bulk ReNiO$_3$ (except Re = La) is paramagnetic metal (PM) at high temperatures and becomes insulating and antiferromagnetically ordered at $T_{\text{MI}}$ and $T_{\text{N}}$, respectively. In particular, a characteristic behavior of the AFM ordering is reflected in the presence of thermal hysteresis in d.c. transport around $T_{\text{N}}$ when $T_{\text{N}}$ is nearby or coincide with $T_{\text{MI}}$. The hysteretic behavior is suppressed or disappear as $T_{\text{N}} \ll T_{\text{MI}}$. We deduced the phase behavior of the NdNiO$_3$ ultrathin films from the resistivity data by identifying two important temperatures. The first one is the resistivity minimum temperature $T^{**}$ which marks the position of $\frac{d\rho}{dT}|_{T=T^{**}} = 0$ signifying the emergence of a charge excitation gap of the MIT. The other characteristic temperature scale is deduced from the resistivity inflection point ($\frac{d\ln\rho}{d(1/T)}|_{\max}$) in the thermal hysteric regime denoted as $T_{\text{N}}$ in Fig.2(a) of the main text. An example is shown in Fig.2(a). The evolution of the transitions controlled by heterointerface constraint is discussed in the main text. From Fig.1, one can also see that the size of the thermal hysteresis (i.e. thermal coercivity) is rapidly decreasing as $\varepsilon$ decreases from +2.7% toward zero, and completely disappears at $\varepsilon = +0.3\%$. This behavior is consistent with that of $T_{\text{N}}$ and further quantified in Fig.2(b). The presence/absence of the $E'$-type AFM ordering is confirmed by measuring the magnetic reflection intensity at $k = (1/4,1/4,1/4)$ by resonant X-ray diffraction (see main text). The 5/3 power-law behavior at $\varepsilon = -2.9\%$ is demonstrated in the fit-free analysis shown in Fig.2(c).

Representative oxygen K-edge x-ray absorption spectrum is shown in a wide energy range in Fig.2(d). The first peak at about 528 eV corresponds to the Ni 3$d$-hybridized states. The second one at about 535 eV comes from states hybridized with the 4$f$ orbitals of Nd and other rare earth elements if contained in the substrates. The features at 540 eV or above mainly originate from bands formed with Ni 4$sp$ states and other metal ions in the substrates.

## II. THEORETICAL MODEL OF NON-FERMI LIQUID PHYSICS

Compared to bulk NdNiO$_3$, one of the main results to emerge in our study is the presence of new electronic phases in the heterointerfacial structure, namely, the paramagnetic insulating (PI) regime, and the emergence of an extended non-Fermi liquid (NFL) regime that appears under compressive strain. As the paramagnetic insulator is featureless (and possibly contains a quantum spin liquid (SL) at zero temperature and small $\varepsilon$), we focus here on the physics likely underpinning the non-Fermi liquid behavior.

The non-Fermi liquid behavior is observed in electrical transport and is shown in Fig.2(d) in the main text. In a Fermi liquid, phase-space arguments applied to low-energy electron-electron scattering around the Fermi energy leads generically to an electrical resistivity that scales as the second power of the temperature, $\rho(T) \propto T^2$. Resistivity that scales as $\rho(T) \propto T^\alpha$ for $\alpha \neq 2$ is considered non-Fermi liquid. As shown in Fig. 2(a) in the main text, for $\varepsilon < 0$, we find $\alpha \neq 2$. In particular, for $-3 \lesssim \varepsilon \lesssim 0$, $\alpha = 4/3$ for $T < T'$, and $\alpha = 1$ for $T > T'$. While for $\varepsilon \lesssim -3$, $\alpha = 5/3$ for $T < T'$, and $\alpha = 1$ for $T > T'$.

The non-Fermi liquid transport exponents imply the presence of degrees of freedom beyond the electrons themselves and are often associated with quantum critical behavior. Thus, common band structure methods, such as density functional theory (DFT), are not able to entirely capture this physics. Nevertheless, DFT is believed to provide a good approximation in many cases to the underlying single-particle physics that interacts with additional degrees of freedom that may be critical. In the following supplementary sections we argue that the main features of the non-Fermi liquid physics we report here can be successfully accounted for by a multi-band model in the presence of zero momentum ($\vec{q} = 0$) critical fluctuations.

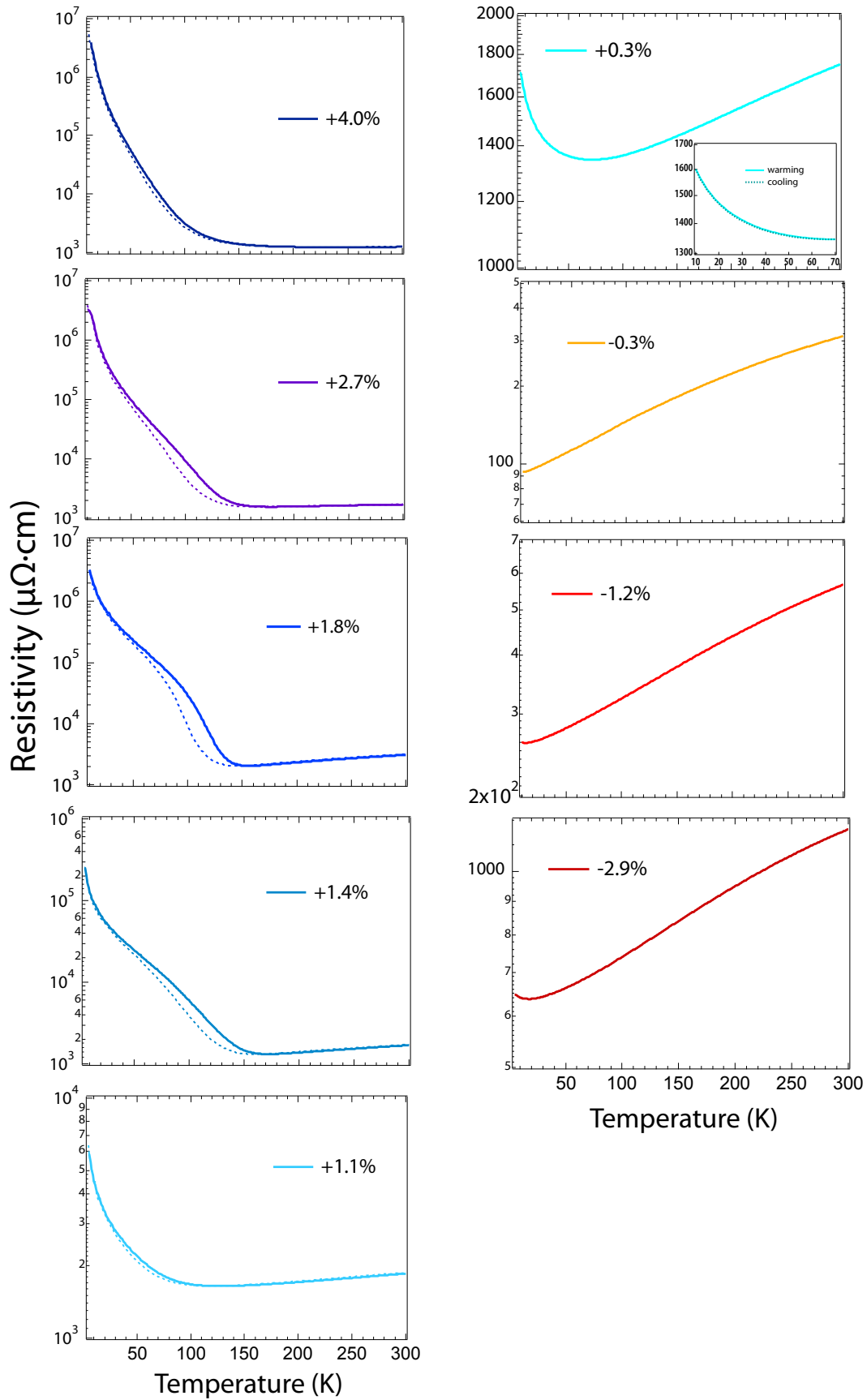

**Figure 1** | Resistivity versus temperature for all lattice mismatches. Dashed lines represent cooling data for samples that have thermal hysteresis. Cooling data at $\varepsilon = +0.3\%$ is included in the inset to demonstrate the absence of hysteresis.

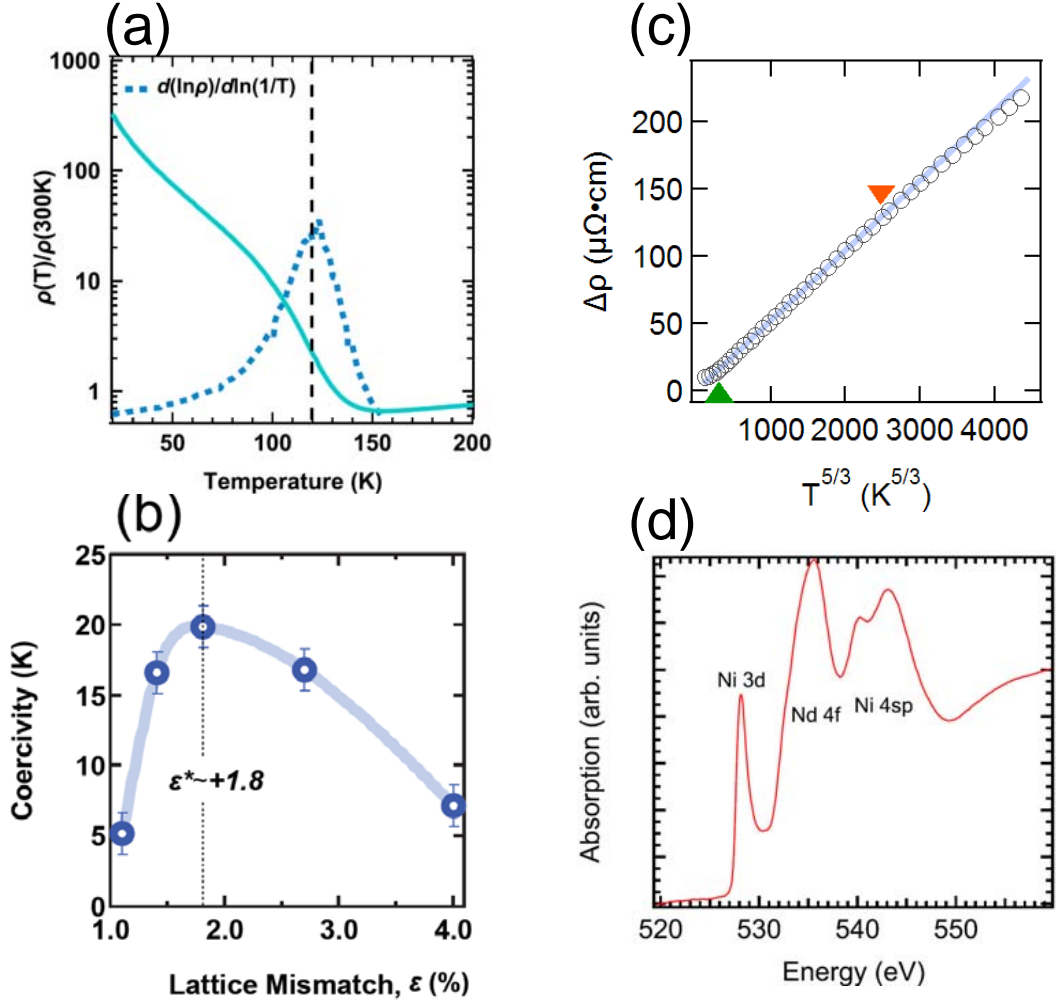

**Figure 2** | a) An example of the resistivity inflection point $T_N$ derived from $(\frac{d\ln\rho}{d(1/T)}|_{\max})$ on the insulating side of the strain ($\varepsilon > 0$). b) The thermal hysteresis coercivity vs strain. c) Fit-free analysis of the $T^{5/3}$ power-law behavior for $\varepsilon = -2.9$. The upper red triangle signals the upper temperature limit $T'$ where the resistivity crosses over to a linear-$T$ behavior. The lower green triangle indicates the lower temperature where the temperature-dependence starts to deviate. d) Oxygen K-edge x-ray absorption spectrum at $\varepsilon = +2.7\%$

### A. Density Functional Theory Results

We compute the electronic band structure of tetragonal NdNiO$_3$ using density functional theory (DFT) with the projector-augmented wave (PAW) formalism[1] as implemented in the Vienna *Ab initio* Simulation Package (VASP)[2,3]. We used the Perdew-Burke-Ernzerhof (PBE) generalized gradient approximation (GGA)[4] plus Hubbard $U$ method[5] (PBE+$U$). We chose the spherically averaged form of the rotationally invariant effective $U$ parameter introduced by Dudarev *et al.*[6] with a $U_{\rm eff} = 4.0$ eV (hereafter, $U$) on the correlated Ni $3d$ orbitals. We impose $A$-type antiferromagnetic (AFM) order as an approximation to the complex $E'$ AFM spin configuration found in the bulk equilibrium phase. The Brillouin zone (BZ) is sampled with a minimum of $9 \times 9 \times 9$ $k$-point mesh, and reciprocal space integrations performed with Gaussian smearing of width 5 meV.

The atomic structure is constrained to have $P4/mmm$ crystallographic symmetry, and we relax the free lattice constants $a$ and $c$ until the cell stresses are zero. For the analysis here, we omit structural distortions due to both rotations of octahedra and breathing distortions to evaluate the electronic structure and band dispersions in a pseudomorphically grown film. The role of octahedral rotations is known to reduce the electronic bandwidth while also doubling the size of the unit cell, leading to BZ effects that otherwise would complicate the analysis here. Note that while the breathing distortion is known to drive a metal–insulator transition at the bulk level, we omit it here so we

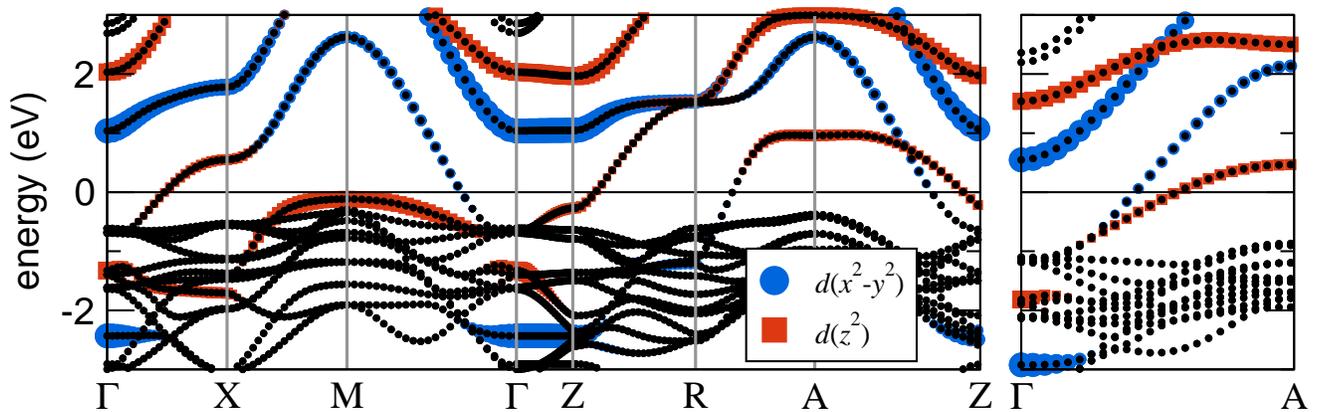

**Figure 3** | Electronic band structure plot for LNO along high-symmetry directions at $\sim 0\%$ strain. The orbital decomposed Ni $d_{z^2}$ (red, light grey) and $d_{x^2-y^2}$ (blue, dark grey) electronic structure for antiferromagnetic NdNiO$_3$. The symbols' size correspond to the magnitude of the Bloch states projected onto each atomic orbital, and black bands close to the Fermi energy are predominantly of oxygen $p$-character. (right) Low-energy electronic structure along the path from the zone-center to zone-corner.

can estimate the band effective masses.

Fig. 3 shows the Ni $e_g$ orbital-resolved electronic band structure throughout the tetragonal BZ using the "fat-bands" method,[7] which allows us to distinguish the different orbital character of the Bloch states near the Fermi level. Consistent with the ionic picture of low-spin Ni$^{3+}$, the states at $E_F$ are primarily of Ni $d_{z^2}$ and $d_{x^2-y^2}$ symmetry with partial occupation of both spin channels. We find spin moments of 0.97 $\mu_B$ on the Ni sites in this configuration with a small amount of spin density located on the oxygen ligands due to itinerancy.

### B. Two Band Model: Classical Approach

The key result from the DFT calculations given above (that should be quite robust, even in the presence of more complicated crystal distortions) is that there are multiple partially occupied bands with rather different effective masses. For example, consider the electron-like dispersion around the $\Gamma$ point and the hole-like dispersion around the A point in Fig.3 of the supplementary information. As was argued in a recent work by Maslov, Yudson, and Chubukov[8], two more bands with different effective masses are key ingredients to stabilize non-Fermi liquid behavior when quantum critical fluctuations are also present in the heavier of the two bands. To motive this result, we first consider a minimal two band model and then later generalize to multiple bands.

#### 1. Two electron bands

The classical equations of motion for two set of carries $s, d$ are,

$$\frac{d\vec{P}_s}{dt} = -e_s \vec{E} - \frac{\vec{P}_s}{\tau_s} - \frac{m_d}{\tau_{sd}}\left(\frac{\vec{P}_s}{m_s} - \frac{\vec{P}_d}{m_d}\right), \tag{1}$$

$$\frac{d\vec{P}_d}{dt} = -e_d \vec{E} - \frac{\vec{P}_d}{\tau_d} - \frac{m_s}{\tau_{sd}}\left(\frac{\vec{P}_d}{m_d} - \frac{\vec{P}_s}{m_s}\right), \tag{2}$$

where $\vec{P}$ is momentum, $e_s$ is the charge of $s$-carriers, $e_d$ the charge of the $d$-carriers, $m_s, m_d$ the respective masses of the carriers, $\vec{E}$ the applied electric field, $\tau_s, \tau_d$ the respective scattering times of the carriers on impurities, and $\tau_{sd}$ the inter-band scattering time. In these equations, the third term on the right-hand side accounts for the friction between carriers, while the first two terms are standard for a single-band model.

Solving these equations and taking the limit $t \to \infty$, we obtain the following expressions for the momenta. We also take $e_s = e_d = e$ for bands of like charge (two electrons, for example),

$$\vec{P}_s = -\frac{\frac{1}{\tau_d} + \frac{1}{\tau_{sd}}(1 + \frac{m_s}{m_d})}{\frac{1}{\tau_s \tau_d} + \frac{1}{\tau_{sd}}(\frac{1}{\tau_s}\frac{m_s}{m_d} + \frac{1}{\tau_d}\frac{m_d}{m_s})} \vec{E}, \quad (3)$$

$$\vec{P}_d = -\frac{\frac{1}{\tau_s} + \frac{1}{\tau_{sd}}(1 + \frac{m_d}{m_s})}{\frac{1}{\tau_s \tau_d} + \frac{1}{\tau_{sd}}(\frac{1}{\tau_s}\frac{m_s}{m_d} + \frac{1}{\tau_d}\frac{m_d}{m_s})} \vec{E}, \quad (4)$$

which leads to the conductivity for each carrier,

$$\sigma_s = \frac{n_s e^2}{m_s} \frac{\frac{1}{\tau_d} + \frac{1}{\tau_{sd}}(1 + \frac{m_s}{m_d})}{\frac{1}{\tau_s \tau_d} + \frac{1}{\tau_{sd}}(\frac{1}{\tau_s}\frac{m_s}{m_d} + \frac{1}{\tau_d}\frac{m_d}{m_s})}, \quad (5)$$

$$\sigma_d = \frac{n_d e^2}{m_d} \frac{\frac{1}{\tau_s} + \frac{1}{\tau_{sd}}(1 + \frac{m_d}{m_s})}{\frac{1}{\tau_s \tau_d} + \frac{1}{\tau_{sd}}(\frac{1}{\tau_s}\frac{m_s}{m_d} + \frac{1}{\tau_d}\frac{m_d}{m_s})}, \quad (6)$$

where $n_{s,d}$ is the electron density for each band. The total resistivity for two electron bands is

$$\rho_{ee} = (\sigma_s + \sigma_d)^{-1}. \quad (7)$$

If one further assumes 2D parabolic bands, the Fermi energy is described in terms of masses and electron density as

$$\epsilon_F = \pi \hbar^2 \frac{n_s}{m_s} = \pi \hbar^2 \frac{n_d}{m_d}, \quad (8)$$

so that the total resistivity for two electron bands $\rho_{ee} = (\sigma_s + \sigma_d)^{-1}$ becomes

$$\rho_{ee} = \frac{\pi \hbar^2}{e^2 \epsilon_F} \frac{\frac{1}{\tau_s \tau_d} + \frac{1}{\tau_{sd}}(\frac{1}{\tau_s}\frac{m_s}{m_d} + \frac{1}{\tau_d}\frac{m_d}{m_s})}{\frac{1}{\tau_s} + \frac{1}{\tau_d} + \frac{1}{\tau_{sd}}(2 + \frac{m_s}{m_d} + \frac{m_d}{m_s})}. \quad (9)$$

This relation is identical to Eq.(6) in Ref.[8]. Moreover, in 3D, or for non-parabolic bands, the numerator is modified as the relation in Eq.(8) no longer holds. The key feature of Eq.(9) is that the dominant temperature dependence of $\rho_{ee}$ enters through $\tau_{sd}$ as the impurity scattering times $\tau_s, \tau_d$ are only weakly temperature dependent and can be taken as a constant to a good approximation when quantum criticality is present.[8]

The key result here is that for a window of temperatures in 2D $\tau_{sd} \propto T^{-4/3}$ with $\vec{q} = 0$ critical fluctuations, while for 3D, $\tau_{sd} \propto T^{-5/3}$. This temperature dependence leads to $\rho_{ee} \propto T^{4/3}$ in 2D and $\rho_{ee} \propto T^{5/3}$ in 3D,[8] which are the exponents we measure in our experiments. Moreover, a more detailed calculation shows (see following sections) that ignoring the "vertex corrections" at higher temperatures will lead to $\rho_{ee} \propto T$ in 3D, also in excellent agreement with our experimental results.

### C. Three Band Model

The equation of motion is generalized to a multiple band model (here we consider three) with their own masses, relaxations, and inter-band scatterings:

$$\frac{d\vec{P}_i}{dt} = -e_i \vec{E} - \frac{\vec{P}_i}{\tau_i} - \sum_{j \neq i} \frac{m_j}{\tau_{ij}} \left( \frac{\vec{P}_i}{m_i} - \frac{\vec{P}_j}{m_j} \right). \quad (10)$$

Here we assume that all three bands are of electron type, but with different masses. Specifically, we assume one of the bands is very massive, i.e. $m_3 \gg m_1, m_2$, implying $\tau_3 \ll \tau_1, \tau_2$. We assume band 3 is close to criticality while the bands 1 and 2 are off-criticality. To make things simpler we also assume the scattering between bands 1 and 2 is not effective, i.e. $\tau_{12} \to \infty$, which is consistent with these bands being off-criticality. We also take $\tau_{13} = \tau_{23} = \tau$. Performing the same steps as above, we arrive at

$$\sigma(T) = \frac{e^2 \epsilon_F}{\pi \hbar^2} \frac{\tau_1 + \tau_2}{m_1 \tau_2 + m_2 \tau_1} \frac{m_1 m_2}{m_3} \tau, \quad (11)$$

$$\rho(T) = \frac{\pi \hbar^2}{e^2 \epsilon_F} \frac{m_1 \tau_2 + m_2 \tau_1}{\tau_1 + \tau_2} \frac{m_3}{m_1 m_2} \frac{1}{\tau}. \quad (12)$$

The temperatures setting the upper and lower limits of the non-Fermi liquid behaviors $\rho(T) \propto T^{4/3}$ in 2D and $\rho(T) \propto T^{5/3}$ in 3D are determined from[8]

$$\frac{1}{\tau(T_1')}\left(\frac{\tau_3}{\tau_1}\frac{m_3}{m_2} + \frac{\tau_3}{\tau_2}\frac{m_3}{m_1} + \frac{m_3}{m_2} + \frac{m_3}{m_1}\right)\frac{\tau_1\tau_2}{\tau_1+\tau_2} = 1 \Rightarrow \frac{1}{\tau(T_1')} = \frac{1}{\tau_3}\frac{1}{m_3}\frac{\tau_1+\tau_2}{\frac{\tau_1}{m_1}+\frac{\tau_2}{m_2}+\frac{\tau_1\tau_2}{\tau_3}\left(\frac{1}{m_1}+\frac{1}{m_2}\right)}, \quad (13)$$

$$\frac{1}{\tau(T_2')}\left(\tau_2\frac{m_3}{m_2} + \tau_1\frac{m_3}{m_1}\right) = 1 \Rightarrow \frac{1}{\tau(T_2')} = \frac{1}{\frac{\tau_1}{m_1}+\frac{\tau_2}{m_2}}\frac{1}{m_3}, \quad (14)$$

which implies that

$$\frac{\tau(T_2')}{\tau(T_1')} \gg 1 \Rightarrow T_1' \gg T_2'. \quad (15)$$

Thus, the scaling regime is $T_2' \ll T \ll T_1'$, where the resistivity goes as $\rho(T) \propto \frac{1}{\tau(T)}$.

We are interesting in how the temperature window of scaling is extended or contracted as the number of bands are increased. To compare with a two band model, we reduce the definitions of characteristic temperatures in Eq.(13) to a two band model by simply dropping band 2. We have,

$$\frac{1}{\tau(T_1)} = \frac{1}{\tau_3}\frac{m_1}{m_3}, \quad (16)$$

$$\frac{1}{\tau(T_2)} = \frac{1}{\tau_1}\frac{m_1}{m_3}. \quad (17)$$

Hence we conclude this window extends as

$$T_2' < T_2 \ll T \ll T_1' < T_1. \quad (18)$$

So, we can see that as the number of bands increase the lower and upper limits of the window is pushed down to lower temperatures. This is seen in Figs. 4 and 5 of the supplementary information.

### D. Boltzmann Equations in the Presence of Impurities

We consider the inter-band scattering via interactions between electrons mediated with critical fluctuations. This was carried out in Ref.[8]. The corresponding coupled Boltzmann equations are,

$$ev_i(k_i) \cdot En_i' = -\sum_{j \neq i} I_{ee}^{ij}[f_i, f_j] - \frac{f_i(k_i) - n_i}{\tau_i}, \quad (19)$$

where $n$ is the undisturbed Fermi distribution, while $f$ is the disturbed one. We also used the relaxation time approximation (relaxation time $\tau_i$) to include the scattering off impurities. $I_{ee}^{ij}$ is the collision integral describing scattering of electrons of the $i$'th band from $j$'th band, and is expressed as,

$$I_{ee}^{ij}[f_i, f_j] = \int \frac{d^2k_j}{(2\pi)^2} \int \frac{d^2k_i'}{(2\pi)^2} \int \frac{d^2k_j'}{(2\pi)^2} W(q,\omega)\Big[f_i(k_i)f_j(k_j)(1-f_i(k_i'))(1-f_j(k_j')) - f_i(k_i')f_j(k_j')(1-f_i(k_i))(1-f_j(k_j))\Big]$$
$$\times \delta(\varepsilon_{k_i} + \varepsilon_{k_j} - \varepsilon_{k_i'} - \varepsilon_{k_j'})\delta(k_i + k_j - k_i' - k_j'), \quad (20)$$

where $W(q,\omega)$ is the interaction matrix. Solving the Boltzmann equations, we obtain the following relation for the resistivity of the system,

$$\rho_{ee} = \frac{\pi\hbar^2}{e^2\varepsilon_F}\frac{\frac{1}{\tau_s\tau_d} + \frac{1}{\tau_{sd}}\left(\frac{1}{\tau_s}\frac{m_s}{m_d} + \frac{1}{\tau_d}\frac{m_d}{m_s}\right)}{\frac{1}{\tau_s} + \frac{1}{\tau_d} + \frac{1}{\tau_{sd}}\left(2 + \frac{m_s}{m_d} + \frac{m_d}{m_s}\right)}, \quad (21)$$

where

$$\frac{1}{\tau_{sd}} = \frac{1}{2T}\frac{\sqrt{m_s m_d}}{\varepsilon_F^2}\int \frac{d\omega}{2\pi}\int \frac{dq}{2\pi} qW(q,\omega)\omega^2 N(\omega)(N(\omega)+1), \quad (22)$$

with $N(\omega)$ the usual boson distribution function.

The bare interaction between $s$ and $d$ electrons $V_{sd}^0$ is mediated by fluctuations,

$$V_{sd}(q,\omega) = \frac{V_{sd}^0}{1 - \chi_{dd}(q,\omega)V_{dd}^0}. \tag{23}$$

Expanding $\chi_{dd}(q,\omega)$,

$$\chi_{dd}(q,\omega) \simeq \chi_0 + cq^2 - ib\frac{\omega}{q}, \tag{24}$$

we can see that the if $V_{dd}^0 \chi_0 \to 1$, the criticality would be associated with dynamical exponent $z=3$. Let's define

$$\chi(q,\omega) = \frac{1}{cq^2 - ib\frac{\omega}{q}}, \tag{25}$$

then we can write the scattering matrix as

$$W(q,\omega) = |V_{sd}|^2 = (\frac{V_{sd}^0}{V_{dd}^0})^2 \frac{1}{b}\frac{q}{\omega}\text{Im}\chi(q,\omega). \tag{26}$$

This allows one to write the scattering rate as

$$\frac{1}{\tau_{sd}} = \frac{1}{2T}\frac{\sqrt{m_s m_d}}{\varepsilon_F^2}(\frac{V_{sd}^0}{V_{dd}^0})^2\frac{1}{b}\int\frac{d\omega}{2\pi}\int\frac{dq}{2\pi}q\frac{\omega}{q}\text{Im}\chi(q,\omega)N(\omega)(N(\omega)+1)\underbrace{q^2}. \tag{27}$$

This relation is identical to the relation for scattering rate close to a ferromagnetic phase transition.[9] Moreover, the term in braces comes from the vertex corrections to render the conductivity finite. The temperature scaling goes as,

$$\frac{1}{\tau_{sd}} \propto T^{4/3} \text{ in 2D}$$
$$\frac{1}{\tau_{sd}} \propto T^{5/3} \text{ in 3D} \tag{28}$$

with vertex corrections, and

$$\frac{1}{\tau_{sd}} \propto T^{2/3} \text{ in 2D}$$
$$\frac{1}{\tau_{sd}} \propto T\ln T \text{ in 3D} \tag{29}$$

without vertex corrections. Generally arguments suggest[10] that vertex corrections can often be ignored at higher temperatures, which we assume is set by the temperature $T'$ in our experiments. Thus, Eq.(28) applies at temperatures below $T'$, and Eq.(29) at temperatures above $T'$.

If we compare the results above with the exponents measured in our experiments, it is tempting to interpret the transport below $T'$ as quasi-two dimensional for $-3 \lesssim \varepsilon \lesssim 0$ and quasi-three dimensional for $\varepsilon \lesssim -3$. Based on the rather general results (mechanism independent), Eq.(28) and Eq.(29), it also appears the transport is quasi-three dimensional above $T'$. For a NdNiO$_3$ to change from effectively two-dimensional to effectively three-dimensional under increasing compressive strain is not unexpected.[11] However, one should check that an estimate of the temperature $T'$ is in rough agreement with the experimentally determined value.

### E. Temperature Dependence of Transport Scattering Rate

In the previous section, one could see that the non-Fermi liquid scaling (with the vertex correction) is achieved within a temperature window–see Eq.(18). In this section we give a rough estimate of the scaling of this window with the number of bands and rough estimates for experimental parameters. A precise determination of the scaling region requires detailed knowledge of the band structure of the model and the exact mechanism of the quantum criticality. We present a simplified case here and consider parabolic bands (though the DFT results predict rather non-trivial Fermi surface shapes). The shape and topology of the different Fermi surfaces will numerically affect the result, but the order of magnitude effects can be determined within this approximation.

Here we use the experimental values for electrical resistivity in the temperature range corresponding to non-Fermi liquid with temperature scaling $T^{4/3}$, to determine what temperature window they predict according to following

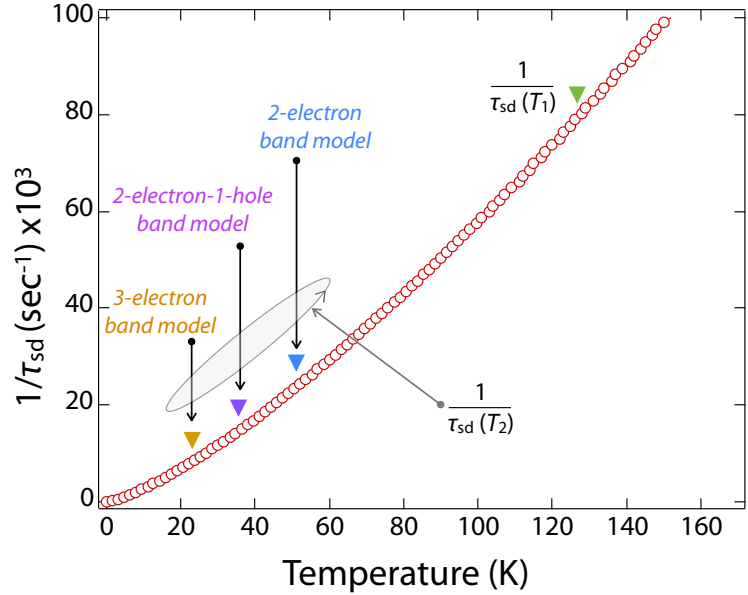

**Figure 4** | (Color online) Inter-band scattering rate ($s^{-1}$) in Eq.(27) vs temperature (in Kelvin). $T_1$ and $T_2$ stand for upper and lower part of scaling regime, where relations in Eq.(30) and Eq.(31) are satisfied. The plot is obtained for $b = c = 1$. Models with different number of bands lead to shifting of scaling regime. Downward arrows: Green arrow denotes upper limit of scaling regime that doesn't change so much with increasing bands, but is suppressed slightly. Above the temperature $T_1$, the resistivity is expected to cross-over to $T \ln T$ [see Eq.(29) and discussion immediately below it], consistent with experiment. Blue, Purple and Brown arrows stand for lower part of scaling regime that are shifted to lower temperature with increasing bands and taking hole-like states as well into account.

formulas derived in Sec.II C for the three band model,

$$\frac{1}{\tau(T_1)} = \frac{1}{\tau_3}\frac{m_1}{m_3},$$
$$\frac{1}{\tau(T_2)} = \frac{1}{\tau_1}\frac{m_1}{m_3}, \tag{30}$$

which can be reduced to the two band model $s-d$ model as derived in Ref[8],

$$\frac{1}{\tau_{sd}(T_1)} = \frac{1}{\tau_s}\frac{m_s}{m_d},$$
$$\frac{1}{\tau_{sd}(T_2)} = \frac{1}{\tau_d}\frac{m_s}{m_d}. \tag{31}$$

Our goal is to see if this approach can yield results in semi-quantitative agreement with experiment. (We find it does in good accord.) We use some typical values for the parametrized susceptibility in Eq.(25): $c = b = 1$ and choose that the inter-band interaction to be very small $\frac{V_{sd}^0}{V_{dd}^0} \sim 10^{-3}$. We numerically calculate Eq.(27) as shown in Fig.4 of the supplemental information.

For a two-band model the upper and lower parts of the $T^{4/3}$ temperature window are exhibited by downward green and blue arrows, respectively. As we showed in Sec.II C, the lower part of window can be pushed down to lower temperatures as the number of bands increases. For example, for the three electron band model (See Sec.II C), it is clear to see that the lower bound is decreased by almost a factor of 2, while the upper bound changes very little. According to our DFT calculations, actual band structure of the system is complex, and exhibits open Fermi surfaces. Perhaps, if these details are taken into account, the lower bound of temperature window will go down even further to the lowest measured temperature.

We also considered another set of parameters ($c = 0.28$ and $b = 5.8$) and the corresponding plot is shown in Fig.5. Although the scaling window shifts to slightly lower temperatures, it still remains within a physical range of temperatures which are not too far away from the experimental values. This indicates that despite the simplifications mentioned above, the model appears to capture the most important underlying physics.

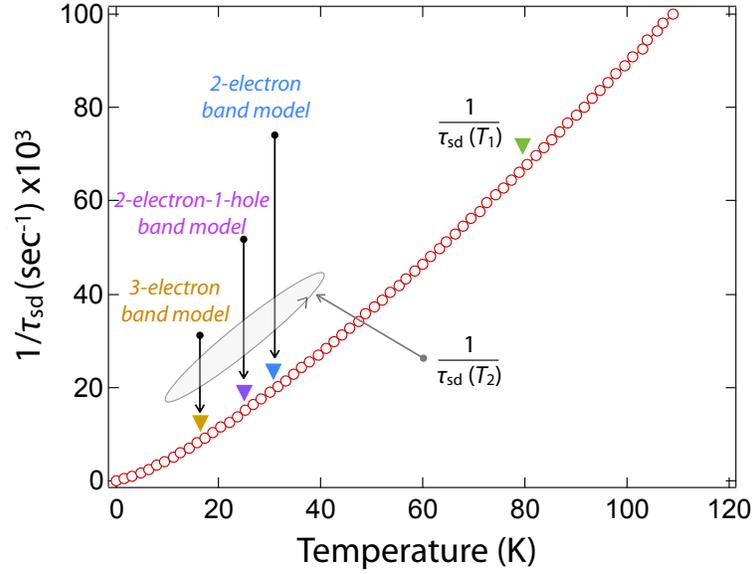

**Figure 5** | (Color online) Same as in Fig.4 but with different set of parameters: $c = 0.28$ and $b = 5.8$

### F. Possible Mechanisms of Criticality: Charge Fluctuations Driven by Interactions and Phonons Modes

In this section we speculate on two possible mechanisms that (in principle) could lead to the exotic scaling behavior observed in our experiment.

#### 1. Fluctuations Driven by Interactions

There is a paramagnetic insulator phase in the phase diagram shown in Fig.2 of main text. This phase is close to a charge disproportionate state with two inequivalent Ni sites.[12] Although magnetic circular dichroism on Ni L-edge measurement exhibits the existence of local magnetism with short range correlations which brings the idea of a spin liquid phase into consideration, we assume that charge disproportionation is a largely relevant description of this phase. More specifically, it would be natural to assume that the non-Fermi liquid phase in proximity to the charge ordered phase inherits some signatures of density fluctuations. Thus, we can assume that on the compressive side the charge density is almost uniform with the presence of long wave-length density fluctuations in it. Then as the compressive strain decreases, the density fluctuations act to give rise to charge ordering on tensile side. Moreover, it seems that the role of oxygen ions is critically important here because of the relevant charge transfer $p-d$ physics in the insulating phase.

In the past, phase separation and associated density fluctuations have been proposed as a possible mechanism for anomalous normal state behavior of cuprates as such fluctuations can induce large scattering amplitude among quasiparticles.[13,14] Both weak[15] and strong[16] coupling limits of the on-site Hubbard interactions indicate the occurrence of a charge instability driven by inter-site interactions between transition ions and oxygens.

A mean-field treatment of the problem shows a transition from a metallic phase to insulating phase. Near the phase transition the density fluctuations give rise to a massless mode. Here the total charge density $n_d + n_p$ and charge transfer density $n_d - n_p$ are strongly fluctuating. These zero modes substantially modify the effective scattering amplitude between quasiparticles via the vertex interaction with bosonic degrees of freedom. These latter modes come from the fluctuations of the fields coupling to the mean field parameters. The corresponding propagators can be described as follows,

$$D(q,\omega) = N^{-1}\Big[2B + \Pi(q,\omega)\Big]^{-1}, \tag{32}$$

where N is the number of fermion flavors in a 1/N expansion, $\Pi(q,\omega)$ is the polarization including vertex interactions, and $B$ describes the spectrum of bosons. The effective interaction between quasiparticles is then given by

$$\Gamma_{kk'}(q,\omega) = -\Lambda(k',-q)D(q,\omega)\Lambda(k,q), \tag{33}$$

where Λ stands for vertex interaction.

The charge instability is given by the diverging of the density-density correlation function leading to
$$\det(2B + \Pi) = 0, \tag{34}$$
that immediately gives rise to a diverging effective interaction. One can show that the spectral density of the total density-density correlation function exhibits a peak at small momentum transfers, *i.e.* $\vec{q} \approx 0$ instability. This mode is damped into the particle-hole continuum, which in turn modifies the self-energy of the quasiparticles thus taking it away from the conventional Fermi-liquid behavior.

### 2. Fluctuations Driven by Phonons

Phonons are also a possible driving force to induce the charge instability in the system. The phonons are coupled to charge density at both transition ions and oxygen sites.[17] The charge instability will always take place by increasing the large bare charge-transfer difference $(\varepsilon_p - \varepsilon_d)/t_{pd}$, where $\varepsilon_{p(d)}$ is the onsite energy and $t_{pd}$ is the transition ion-oxygen hopping. This latter can be easily tuned in our system via the compressive strain. Near the phase separation instability, the anomalous scattering amplitude is identified to be of the following form[14,18]
$$\Gamma(q, \omega) \sim -\frac{1}{Bq^2 - iC\frac{\omega}{q} + D}, \tag{35}$$
where $D = D(\delta - \delta_c)$ measures the closeness to criticality via doping. As clearly seen this effective interaction is singular with dynamical exponent $z = 3$, which in turn strongly modifies the scattering relaxation and transport rates; the singular interaction thus make them to have non-Fermi liquid character. In both the cases above, an enlarged unit cell would naturally lead to $\vec{q}$ wavevector fluctuations.

## III. ACKNOWLEDGEMENT

## REFERENCES

[1] P. E. Blöchl, Phys. Rev. B, **50**, 17953 (1994).
[2] G. Kresse and J. Furthmller, Computational Materials Science, **6**, 15 (1996), ISSN 0927-0256.
[3] G. Kresse and D. Joubert, Phys. Rev. B, **59**, 1758 (1999).
[4] J. P. Perdew, K. Burke, and M. Ernzerhof, Phys. Rev. Lett., **77**, 3865 (1996).
[5] V. I. Anisimov, F. Aryasetiawan, and A. I. Lichtenstein, Journal of Physics: Condensed Matter, **9**, 767 (1997).
[6] S. L. Dudarev, G. A. Botton, S. Y. Savrasov, C. J. Humphreys, and A. P. Sutton, Phys. Rev. B, **57**, 1505 (1998).
[7] O. Jepsen and O. Andersen, Zeitschrift für Physik, **97** (1995).
[8] D. L. Maslov, V. I. Yudson, and A. V. Chubukov, Phys. Rev. Lett., **106**, 106403 (2011).
[9] R. P. Smith, M. Sutherland, G. G. Lonzarich, S. S. Saxena, N. Kimura, S. Takashima, M. Nohara, and H. Takagi, Nature, **455**, 1220 (2008).
[10] G. Mahan, *Many-Body Particles* (Kluwer Academic Plenum Publishers).
[11] R. J. Angel, J. Zhao, and N. L. Ross, Phys. Rev. Lett., **95**, 025503 (2005).
[12] J. Chakhalian, J. M. Rondinelli, J. Liu, B. A. Gray, M. Kareev, E. J. Moon, N. Prasai, J. L. Cohn, M. Varela, I. C. Tung, M. J. Bedzyk, S. G. Altendorf, F. Strigari, B. Dabrowski, L. H. Tjeng, P. J. Ryan, and J. W. Freeland, Phys. Rev. Lett., **107**, 116805 (2011).
[13] V. Emery and S. Kivelson, Physica C: Superconductivity, **209**, 597 (1993), ISSN 0921-4534.
[14] C. Castellani, C. Di Castro, and M. Grilli, Phys. Rev. Lett., **75**, 4650 (1995).
[15] Y. Bang, G. Kotliar, C. Castellani, M. Grilli, and R. Raimondi, Phys. Rev. B, **43**, 13724 (1991).
[16] R. Raimondi, C. Castellani, M. Grilli, Y. Bang, and G. Kotliar, Phys. Rev. B, **47**, 3331 (1993).
[17] M. Grilli and C. Castellani, Phys. Rev. B, **50**, 16880 (1994).
[18] F. Becca, M. Tarquini, M. Grilli, and C. Di Castro, Phys. Rev. B, **54**, 12443 (1996).




\newpage
\begin{thebibliography}{99}

\bibitem{Zaanen}
J. Zaanen, G. A. Sawatzky, and J. W. Allen, Phys. Rev. Lett. \textbf{55}, 418 (1985).

\bibitem{Chakhalian2012}
J. Chakhalian A. J. Millis, and J. Rondinelli, Nature Mater. \textbf{11}, 92 (2012).

\bibitem{Mannhart}
J. Mannhart and D. G. Schlom, Science \textbf{327}, 1607 (2010).

\bibitem{Jian}
Jian Liu et al., Appl. Phys. Lett. \textbf{96}, 233110 (2010).

\bibitem{Supplemental}
see On-line  supplemental  material.

\bibitem{Catalan0}
G. Catalan, Phase Transitions \textbf{81}, 729 (2008).

\bibitem{Medarde1}
M. L. Medarde, J. Phys.: Condens. Matter \textbf{9}, 1679 (1997).

\bibitem{Mizokawa0}
T. Mizokawa et al., Phys. Rev. Lett. \textbf{67}, 1638 (1991)

\bibitem{Sarma}
D. D. Sarma, J. Solid State Chem. \textbf{88}, 45 (1990).

\bibitem{Khomskii}
D. I. Khomskii, Lith. Phys. J. \textbf{37}, 65 (1997); arXiv:cond-mat/0101164v1.

\bibitem{Mizokawa}
T. Mizokawa, D. I. Khomskii, and G. A. Sawatzky, Phys. Rev. B \textbf{61}, 11263 (2000).

\bibitem{Alonso}
J. A. Alonso et al., Phys. Rev. Lett. \textbf{82}, 3871 (1999).

\bibitem{Staub}
U. Staub et al., Phys. Rev. Lett. \textbf{88}, 126402 (2002).

\bibitem{Goodenough}
J. B. Goodenough, J. Solid State Chem. \textbf{127}, 126 (1996).

\bibitem{Doering}
D. Doering et al., Rev. Sci. Instrum. \textbf{82}, 073303 (2011).

\bibitem{Bodenthin}
Y. Bodenthin et al.,
%U. Staub, C. Piamonteze, M. Garca-Fernndez, M. J. Martnez-Lope, and J. A. Alonso,
J. of Phys.: Cond. Matt. \textbf{23}, 036002 (2011).

\bibitem{Balents}
L. Balents,
%``Spin liquids in frustrated magnets",
Nat. \textbf{464}, 199 (2010).

\bibitem{Suter}
A. Suter et al., Phys. Rev. Lett. \textbf{106}, 237003 (2011).

\bibitem{Debye}
K.P. Rajeev, G. V. Shivashankar, and A. K. Raychaudhuri, Solid State Commun. 79, 591 (1991).

\bibitem{Imada}
M. Imada, A. Fujimori, and Y. Tokura, Rev. Mod. Phys. \textbf{70}, 1039 (1998).

\bibitem{Maslov}
D. L. Maslov, V. I. Yudson and A. V. Chubukov,
%``Resistivity of a Non-Galilean Invariant Fermi Liquid near Pomeranchuk Quantum Criticality ",
Phys. Rev. Lett. \textbf{106}, 106403 (2011).

\bibitem{Angel}
R. J. Angel, J. Zhao, and N. L. Ross, Phys. Rev. Lett. {\bf 95}, 025503 (2005).

\bibitem{Zhou0}
J.-S. Zhou, J. B. Goodenough, and B. Dabrowski, Phys. Rev. Lett. \textbf{94}, 226602 (2005).

\bibitem{Sarma02}
D. D. Sarma, N. Shanthi, and P. Mahadevan, Phys. Rev. B \textbf{54}, 1622 (1996).

\bibitem{Merz}
M. Merz et al., Phys. Rev. Lett. \textbf{80}, 5192 (1998).

\bibitem{footnote}
Features at much higher energies, such as the Nd 4$f$, 5$sp$ states, are also shifting in the same manner as the Ni 3d prepeak. These high-energy features are, however, often interfered by spectral contribution from the substrates and not crucial for the low-energy physics. They are hence not shown.

\bibitem{Ohta}
Y. Ohta, T. Tohyama, and S. Maekawa, Phys. Rev. Lett. \textbf{66}, 1228 (1991).

\bibitem{Zaanen1}
J. Zaanen and G. A. Sawatzky, J. Solid State Chem. \textbf{88}, 8 (1990).



%\bibitem{Chou}
%F. C. Chou et al., Phys. Rev. Lett. \textbf{75}, 2204 (1995).







\end{thebibliography}
\end{document}